# Strong-coupling-assisted formation of coherent radiation below the lasing threshold

I. V. Doronin,[1,2,3] A. A. Zyablovsky,[1,2,3] E. S. Andrianov[1,2,3]

[1]Dukhov Research Institute of Automatics (VNIIA), 22 Sushchevskaya, Moscow 127055, Russia

[2]Moscow Institute of Physics and Technology, 9 Institutskiy per., Dolgoprudny 141700, Moscow reg., Russia

[3]Institute for Theoretical and Applied Electromagnetics, 13 Izhorskaya, Moscow 125412, Russia

**Abstract:** The creation of nanoscale lasers that operate above a coherent threshold is a challenging problem. We propose a way to circumvent this issue using systems in which a strong coupling regime is achieved between the light and the active medium. In the strong coupling regime, energy oscillations take place between the EM field in the cavity and the atoms. By applying appropriate time modulation to the pumping, it is possible to control these energy oscillations in such a way that coherence in the laser system appears below the lasing threshold. We show that in this approach, the radiation linewidth is two orders of magnitude smaller than the linewidth of a conventional laser for the same photon number. In addition, the second order coherence function of the output radiation is reduced from two to one before the system reaches a positive population inversion. Our results pave the way for the creation of nanoscale sources of coherent radiation that can operate below the lasing threshold.

## 1. Introduction

The creation of nano-sized sources of coherent radiation is a challenging problem. The minimum possible size of dielectric lasers is restricted by the half-wavelength of the generated light. Using plasmonic nanostructures as laser resonators enables us to overcome the size limitations arising due to diffraction limits [1-3]. However, the sub-wavelength localization of the EM field in the plasmonic lasers results in large ohmic losses. To achieve lasing, it is necessary to overcome these losses, which requires high gain in the active medium and high pumping power. However, high pumping power leads to heating and degradation of the active medium, which negatively impacts the performance of nanolasers [4, 5]. Moreover, the high losses in plasmonic lasers prevent a reduction in the laser linewidth.

On the other hand, a reduction in the size of a plasmonic laser results in strong localization of the EM field in the plasmonic structures, and consequently an increase in the coupling constant between the EM field and the active atoms. If the coupling constant exceeds the dissipation rate, a system moves to a strong coupling regime [6-12]. In this strong coupling regime, the interaction between the EM field and the atoms or molecules results in the formation of hybrid polaritonic states and the appearance of Rabi splitting in the system spectrum [7-11]. This formation of hybrid states leads to changes in the optical [13, 14], electrical [15, 16] and chemical [17-19] properties of the medium, which can be used, for example, to control the chemical reactions [18-20] or to tune the electrical conductivity [15] and work function [16]. Strongly coupled light–matter systems show promise for the development of polariton circuits [21], single-photon switches [22, 23] and all-optical logic elements [13]. In addition, these systems can serve as building blocks for quantum computers [24-26] and detectors [26].

In a laser system with a single-mode cavity, a strong coupling regime is achieved with a negative population inversion in the active medium [27, 28]. In this regime, energy oscillations between the cavity EM field and atoms occur. During oscillations, there are time intervals in which energy flows from the atoms to EM field, resulting in light amplification in the cavity. Notably, this occurs for a negative population inversion of the atoms. Recently, it was demonstrated [27] that time-modulated pumping enables these energy oscillations to be controlled and, as a result, light amplification to be achieved even for a negative population inversion of the active medium [27]. However, light amplification by itself does not guarantee coherence of the emitted light; to achieve coherence in the sense of the first- and second-order coherence functions, it is necessary to overcome incoherent contributions from noise [29, 30]. In this context, obtaining a coherent signal below the lasing threshold by utilizing the properties of the system in the strong-coupling regime is important problem.

In this paper, we consider the coherent properties of a laser system in the strong coupling regime with time-modulated pumping. We show that two peaks appear in the system spectrum at the frequencies determined by Rabi splitting. The linewidths of these peaks narrow with an increase in the pump rate. Surprisingly, the radiation linewidth in this regime is two orders of magnitude smaller than the Schawlow-Townes linewidth of a conventional laser at the same photon number, allowing us to achieve a narrow linewidth for a small number of photons. Notably, this is achieved at negative population inversions, where conventional lasers do not provide coherent radiation. We demonstrate that the second-order coherence function of radiation from the system changes from two to one with an increase in pumping, meaning that the radiation becomes coherent below the lasing threshold. A reduction in the pumping power required to produce coherent light can solve the problems of overheating and degradation of the active medium in nanolasers.

## 2. Model of a single-mode parametric laser without inversion

In this section, we examine the behavior of a system consisting of a single-mode cavity filled with two-level atoms that are subject to periodic variation in pumping. The cavity mode and the atoms form a strongly-coupled system. This can be achieved, for example, in cavities based on photonic [7] or plasmonic [12, 31] structures.

To describe this cavity-atom system, we employ the Maxwell-Bloch equations with noise terms [29, 32]:

$$\frac{da}{dt} = -\left(\gamma_a + i\omega_0\right)a - iN_{at}\Omega_R\sigma + F_a(t) \quad (1)$$

$$\frac{d\sigma}{dt} = -\left(\gamma_\sigma + i\omega_0\right)\sigma + i\Omega_R^* aD + F_\sigma(t) \quad (2)$$

$$\frac{dD}{dt} = \left(\gamma_P - \gamma_D\right) - \left(\gamma_P + \gamma_D\right)D + 2i\left(\Omega_R a^*\sigma - \Omega_R^* a\sigma^*\right) + F_D(t) \quad (3)$$

Here, $a$ is the electric field amplitude in the cavity mode, and $\sigma$ and $D$ are the average polarization and population inversion of the atoms. $\omega_0$ is the transition frequency of the atoms, which coincides with the resonator frequency. $N_{at} = 10^6$ is the number of atoms in the cavity; $\Omega_R$ is the coupling constant between an individual active atom and the cavity; $\gamma_a$ is the loss rate of the cavity; $\gamma_D$ is the longitude relaxation rate of the active atoms; $\gamma_P(t)$ is the pump rate, which varies in time; and $\gamma_\sigma$ is the transverse relaxation rate of the atomic dipole moment, which is determined by other relaxation rates as $\gamma_\sigma(t) = (\gamma_P(t) + \gamma_D)/2 + \gamma_{deph}$, where $\gamma_{deph}$ is the dephasing rate of the atom's polarization. $F_a(t)$, $F_\sigma(t)$, and $F_D(t)$ are the noise terms for the respective variables. These noise terms enable us to take into account the spontaneous emission processes in the laser [29, 33].

The following values of the system parameters are used: $N_{at} = 10^6$, $\Omega_R = 2 \times 10^{-5} \omega_0$, $\gamma_a = 5 \times 10^{-5} \omega_0$, $\gamma_D = 5 \times 10^{-4} \omega_0$, $\gamma_{deph} = 5 \times 10^{-4} \omega_0$. For the noise terms, we use the following correlations properties: $\langle F_a^*(t) F_a(t) \rangle = 0$, $\langle F_D^*(t) F_D(t) \rangle = 0$, $\langle F_\sigma^*(t) F_\sigma(t) \rangle = (\gamma_D + \gamma_P(t) + \gamma_{deph}(D_0 + 1))/2$. Other noise correlators vanish at room temperature [33].

### 3. Results

*Generation of coherent light below the lasing threshold*

It is known that a strong coupling regime can be realized in lasers [27, 28]. Below the lasing threshold, we can linearize Eqs. (1) and (2), based on the assumptions that the population inversion $D$ does not depend on the amplitudes of the EM field and the polarization, and that $D = D_0$. Here, $D_0 = (\gamma_P - \gamma_D)/(\gamma_P + \gamma_D)$ is a stationary population inversion below the lasing threshold. The linearized equations (1) and (2) have two hybrid eigenstates with complex eigenfrequencies $\omega_{1,2} = \omega_0 \pm i\sqrt{(\gamma_\sigma - \gamma_a)^2/4 + \Omega_R^2 N_{at} D_0}$ [27, 28], where the real parts determine the oscillation frequencies and the imaginary parts determine the relaxation rates.

In weak coupling regime, two eigenstates have the same oscillation frequencies and different relaxation rates; in the strong coupling regime, the oscillation frequencies of the laser eigenstates are split, while the relaxation rates coincide. The transition from the strong to the weak coupling regime occurs at a negative population inversion of atoms, when the population inversion $D_0$ equals [27]:

$$D_{EP} = -\frac{(\gamma_a - \gamma_\sigma)^2}{4\Omega_R^2 N_{at}} < 0 \qquad (4)$$

This transition from the weak to strong coupling regimes ($D_0 = D_{EP}$) corresponds to an exceptional point (EP) at which two eigenstates of the laser system are linearly dependent and their eigenfrequencies are equal to each other [27]. Note that the population inversion of atom can only assume values from −1 to 1, and hence if $D_{EP} < -1$, the strong coupling regime cannot arise in the system.

We consider a laser in the strong coupling regime, in which the pump rate is varied over time as $\gamma_P(t) = \bar{\gamma}_P + \delta\gamma_P \sin(\omega_M t)$, where $\bar{\gamma}_P$ is the mean value of the pump rate, and $\omega_M$ and $\delta\gamma_P$ are the modulation frequency and amplitude, respectively. In Ref. [27], it was demonstrated that a periodic variation in the pump rate enables light amplification even for a negative population inversion in the active medium. This amplification is realized when the system is in the strong coupling regime, $\bar{D}_0 = (\bar{\gamma}_P - \gamma_D)/(\bar{\gamma}_P + \gamma_D) < D_{EP}$, and the modulation frequency $\omega_M(\bar{D}_0)$ is equal to the frequency of Rabi splitting, $\omega_M(\bar{D}_0) = \sqrt{\left|(\gamma_\sigma - \gamma_a)^2 + 4\Omega_R^2 N_{at} \bar{D}_0\right|}$ [27].

Light amplification is not sufficient to obtain coherent radiation with a narrow linewidth [34]. To achieve coherence in the sense of the first- and second-order coherence functions, it is necessary to overcome the incoherent contributions from noise [29, 30]. In a conventional laser, this is realized only above the lasing threshold [29]; however, in a strong coupling laser system, radiation can become coherent when the average pump rate $\bar{\gamma}_P < \gamma_D$ and the population inversion of atoms is negative. Numerical simulation of Eqs. (1)–(3) shows that the EM field intensity has a threshold behavior that depends on the modulation amplitude $\delta\gamma_P$ (Fig. 1a). Two peaks in the system spectrum can be observed at frequencies $\omega_0 \pm \omega_M/2$ (Fig. 1b). The linewidth of each peak narrows with an increase in $\delta\gamma_P$ (Fig. 1a), and the radiation becomes coherent. For this reason, we will refer to this system as a new type of laser: a strong coupling laser (SCL or SC laser).

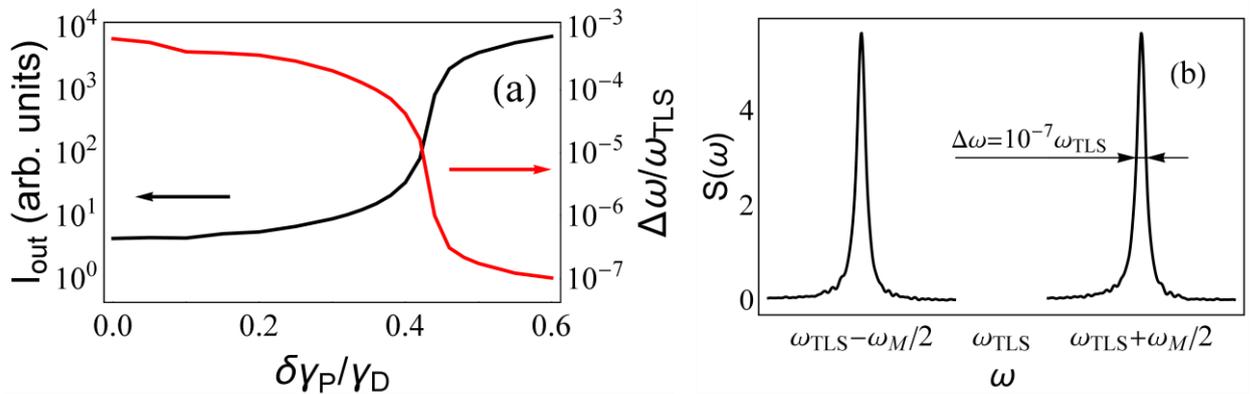

Figure 1. (a) Dependence of the EM field intensity (black solid line) and the radiation linewidth (red dashed line) on the modulation amplitude $\delta\gamma_P$, when the pump rate $\gamma_P(t) = \bar{\gamma}_P + \delta\gamma_P \sin(\omega_M t)$, the average pump rate $\bar{\gamma}_P = 0.91\gamma_D$ and the modulation frequency $\omega_M = 8.6 \times 10^{-2} \omega_{TLS}$ are fixed; (b) system spectrum for the same parameters as in Fig. 1a.

*Comparison of the coherence characteristics of radiation from a strong coupling laser and a conventional laser*

For a more detailed study of the radiation properties of an SCL, we compare the coherence characteristics of an SCL and a conventional laser with continuous wave (CW) pumping for the same parameters of the cavity and active medium. In a CW laser, the pump rate does not depend on time, i.e. $\gamma_P(t) = \bar{\gamma}_P$, whereas in an SCL, the pump rate varies over time as $\gamma_P(t) = \bar{\gamma}_P + \delta\gamma_P \sin(\omega_M t)$. To compare an SCL with a CW laser, we fix the modulation depth, $\delta\gamma_P = 0.6\gamma_D$ and change the average pump rate $\bar{\gamma}_P$ and modulation frequency $\omega_M(\bar{D}_0) = \sqrt{\left|(\gamma_\sigma - \gamma_a)^2/4 + \Omega_R^2 N_{at}\bar{D}_0\right|}$. We examine the dependences of the output, linewidth and second-order coherence function of radiation on the average pump rate $\bar{\gamma}_P$. In this case, the average energy supplied by the pump to both lasers is the same.

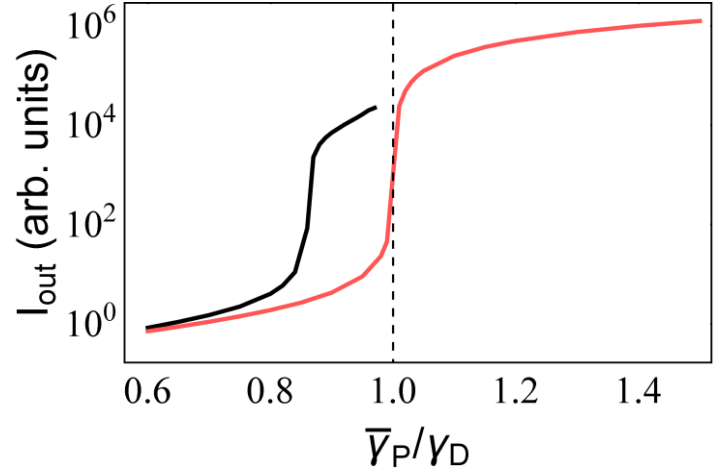

Figure 2. Number of photons in an SCL (black curve) and a conventional laser (red curve) versus the average pump rate $\bar{\gamma}_P$. The vertical dashed line shows $\bar{\gamma}_P$, for which $\bar{D}_0 = D_{EP}$. When $\bar{D}_0 > D_{EP}$, there is no splitting between the eigenfrequencies ($\omega_M(\bar{D}_0) = 0$), and SCL operates as a conventional laser.

Both the CW laser and the SCL exhibit a sharp increase in the average number of photons (and thus in the output power) at a certain threshold pump rate (Fig. 2). However, the threshold for the SCL is considerably smaller than that for the CW laser. The radiation linewidth and coherence function $g^{(2)}(0)$ abruptly decrease for both lasers above their respective thresholds (Fig. 3). Since the threshold for the SCL is lower than that for the CW laser, the SC laser reaches same values of coherence at a lower value of the average pump rate (Fig. 3).

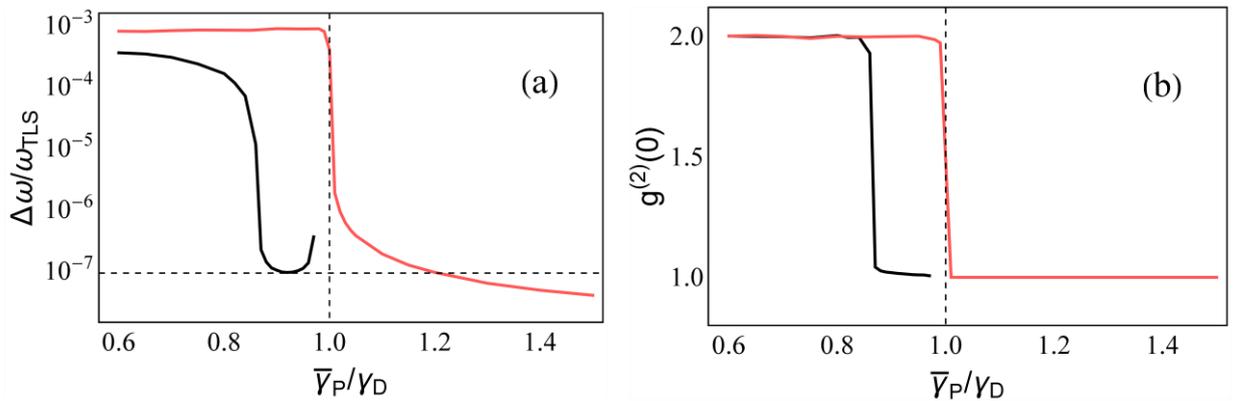

Figure 3. (a) Spectral width of SCL emission (black curve) and conventional laser emission (red curve) versus the average pump rate $\bar{\gamma}_P$; (b) second-order autocorrelation function of SCL emission (black curve) and conventional laser emission (red curve) versus the average pump rate $\bar{\gamma}_P$. The vertical dashed lines show $\bar{\gamma}_P$, for which $\bar{D}_0 = D_{EP}$. When $\bar{D}_0 > D_{EP}$, there is no splitting between the eigenfrequencies ($\omega_M(\bar{D}_0) = 0$), and SCL operates as a conventional laser.

Moreover, the SC laser displays unusual coherence properties. The radiation linewidth of a conventional laser has a lower limit defined by the Schawlow-Townes relation, in which the minimal linewidth $\Delta\omega$ is inversely proportional to the average number of photons $\langle n \rangle$ [29]:

$$\Delta\omega = \frac{A_{CW}}{\langle n \rangle} \quad (5)$$

where $A_{CW}$ is a proportionality factor for the CW laser [32]. Hence, to achieve a narrow radiation linewidth, a conventional laser must operate well above the lasing threshold when $\langle n \rangle \gg 1$.

In the SCL, however, the radiation linewidths are two orders smaller than the Schawlow-Townes linewidth for the same photon number. Indeed, when the average pump rate $\bar{\gamma}_P$ exceeds the threshold for the CW laser, the number of photons in the CW laser is about two orders of magnitude greater than that for the SCL (Fig. 2). At the same time, the radiation linewidth of the CW laser is about the same as that for the SCL (Fig. 3a). Thus, the SCL becomes a generator of coherent radiation at a smaller average value of the pump rate and a much smaller number of photons than conventional CW lasers.

### 4. Discussion

We have shown that an SCL becomes a generator of coherent radiation for much smaller numbers of photons than for a CW laser. This is because spontaneous emission affects the generation of coherent light differently in the cases of a CW laser and an SCL.

In a CW laser, above the lasing threshold, the spontaneous emission of atoms leads to excitation of the EM field oscillations, whose phase is not consistent with the current phase of the generated EM field. Consequently, the phase of the EM field changes chaotically over time, resulting in a finite linewidth of laser radiation. The linewidth is determined by the ratio of the intensity of the EM field excited by the spontaneous emission to the total intensity of the EM field in the cavity [29].

In an SCL, the two hybrid states of the cavity electric field and matter polarization participate in the generation of coherent radiation. The electric field has the form:

$$a(t) = a_1(t) \exp(i\varphi_1(t))\exp(-i\omega_1 t) + a_2(t) \exp(i\varphi_2(t))\exp(-i\omega_2 t) \quad (6)$$

Here, $\omega_{1,2}$ are the oscillation frequencies of the two hybrid states, and $a_{1,2}(t)$ and $\phi_{1,2}(t)$ are the slowly varying amplitudes and phases of these two hybrid states, respectively. Spontaneous emission (noise) results in fluctuation in the phases $\phi_{1,2}(t)$, which is the reason for the nonzero radiation linewidths of both peaks. Eq. (6) can be rewritten as:

$$a(t) = \exp(i\phi)\left[a_1 \exp(i\Delta\phi/2) \exp(-i\omega_1 t) + a_2 \exp(-i\Delta\phi/2) \exp(-i\omega_2 t)\right] \quad (7)$$

where $\phi(t) = (\phi_1(t) + \phi_2(t))/2$ is the common phase of states and $\Delta\phi(t) = \phi_1(t) - \phi_2(t)$ is the relative phase between the states.

Modulation of pumping causes a periodic change in the relative phase between the states, forming an effective potential for $\Delta\phi$. The equations for a strongly coupled laser system with modulated pumping are reduced to the equation for a parametric oscillator [27] (Mathieu's equation), which is equivalent to the Helmholtz equation with a periodic potential. This periodic potential creates a correlation between the phases of the two hybrid states, and can result in the suppression of diffusion of $\Delta\phi$. This type of suppression mechanism is realized in a holographic laser [29, 35-37] where suppression of diffusion of the relative phase between the states leads to a decrease in the linewidths. The diffusion of the common phase $\phi(t)$ in holographic laser prevents the linewidths from being reduced to zero.

In the case of an SC laser, in addition to suppression of fluctuations in the relative phase, $\Delta\phi$, suppression of the common phase, $\phi$, takes place. The reason for this is the strong non-orthogonality of the eigenstates near the EP, due to which the spontaneous emission of atoms predominantly leads to fluctuations in the relative phase, $\Delta\phi$, rather than the common phase, $\phi$ (see Appendix).

Thus, in an SC laser, modulation of pumping leads to suppression of the fluctuations in the relative phase, $\Delta\phi$, when operating near the EP, resulting in suppression of the fluctuations in the common phase, $\phi$. As a result, significant narrowing of both peak linewidths takes place in an SCL. This narrowing can be observed from numerical simulation of Eqs. (1)–(3).

Note that the mechanisms resulting in suppression of phase fluctuations in the SCL have no equivalent in a CW laser. For this reason, a larger number of photons are required to achieve light coherence in a CW laser.

**Conclusion**

Localization of light at the nanoscale can result in a regime of strong coupling between the light and the active atoms. In this regime, hybrid states of light and active atoms appear, and energy oscillations occur between the EM field in the cavity and the atoms. Time-modulated pumping can be used to control these energy oscillations and to achieve light amplification at a negative population inversion of the active medium. In this paper, we demonstrate that the light amplification caused by time-modulated pumping can result in the generation of coherent light in the laser system below the lasing threshold. In this regime, the spectrum for the laser system has two peaks at frequencies determined by Rabi splitting. The linewidths of these peaks are two orders of magnitude smaller than the Schawlow-Townes linewidth for a conventional laser with the same photon number. In this way, it becomes possible to achieve a narrow linewidth for a small number of photons and sub-threshold pumping. The generation of coherent light in the sub-threshold regime paves the way for the creation of nanoscale lasers, which are less prone to overheating and degradation.

**Appendix**

The spontaneous emission of atoms excites oscillations in both the eigenstates of the laser system. The contribution of a single act of spontaneous emission can be represented as $\boldsymbol{\delta f} = \delta f_1 \mathbf{e}_1 + \delta f_2 \mathbf{e}_2$, where $\mathbf{e}_{1,2}$ are the eigenstates of the laser system and $\delta f_{1,2}$ are expansion coefficients determined by the fluctuation.

The same contribution can be written as $\boldsymbol{\delta f} = b_1 \mathbf{e}_1 + b_\perp \mathbf{e}_\perp$, where $\mathbf{e}_\perp$ is a state orthogonal to $\mathbf{e}_1$, i.e., the scalar product $\langle \mathbf{e}_1 | \mathbf{e}_\perp \rangle = 0$. The eigenstate $\mathbf{e}_2$ is written as $\mathbf{e}_2 = \langle \mathbf{e}_2 | \mathbf{e}_1 \rangle \mathbf{e}_1 + \langle \mathbf{e}_2 | \mathbf{e}_\perp \rangle \mathbf{e}_\perp$ and $\boldsymbol{\delta f}$ takes form

$$\boldsymbol{\delta f} = \left( b_1 - \frac{b_\perp}{\langle \mathbf{e}_2 | \mathbf{e}_\perp \rangle} \langle \mathbf{e}_2 | \mathbf{e}_1 \rangle \right) \mathbf{e}_1 + \frac{b_\perp}{\langle \mathbf{e}_2 | \mathbf{e}_\perp \rangle} \mathbf{e}_2 \quad (A1)$$

If $|b_\perp \langle \mathbf{e}_2 | \mathbf{e}_1 \rangle / \langle \mathbf{e}_2 | \mathbf{e}_\perp \rangle| \gg |b_1|$ and $\arg\langle \mathbf{e}_2 | \mathbf{e}_1 \rangle \approx 2\pi n, n = 0,1,...$, then the expansion coefficients of $\boldsymbol{\delta f}$ are approximately equal in modulus and have opposite signs [38], i.e., $\delta f_1 = -\delta f_2 = \delta f$. Near the exceptional point $\mathbf{e}_1$ almost coincides with $\mathbf{e}_2$ (i.e., $|\langle \mathbf{e}_1 | \mathbf{e}_2 \rangle| \simeq 1$ and $|\arg\langle \mathbf{e}_2 | \mathbf{e}_1 \rangle| \ll 1$) and $|\langle \mathbf{e}_2 | \mathbf{e}_\perp \rangle| \ll 1$; and the conditions $|b_\perp \langle \mathbf{e}_2 | \mathbf{e}_1 \rangle / \langle \mathbf{e}_2 | \mathbf{e}_\perp \rangle| \gg |b_1|$ and $\arg\langle \mathbf{e}_2 | \mathbf{e}_1 \rangle \approx 2\pi n, n = 0,1,...$ are satisfied for most possible values of $\boldsymbol{\delta f}$.

Thus, near the EP, the expansion coefficients in $\delta f_1$ and $\delta f_2$ *for almost all* $\boldsymbol{\delta f}$ are approximately equal in modulus and have opposite signs [38], i.e., $\delta f_1 = -\delta f_2 = \delta f$. In this case, for an electric field and taking into account the contribution from spontaneous emission, we obtain:

$$a(t) = \exp(i\phi) \begin{bmatrix} (a_1 \exp(i\Delta\phi/2) + \delta a_1 \exp(-i\phi)) \exp(-i\omega_1 t) + \\ (a_2 \exp(-i\Delta\phi/2) - \delta a_2 \exp(-i\phi)) \exp(-i\omega_2 t) \end{bmatrix} \quad (A2)$$

where $t$ is counted from the moment of spontaneous emission; $\delta a_{1,2} = \delta f \cdot e_{1,2}^{(a)}$, where $e_{1,2}^{(a)}$ are the components of eigenstates $\mathbf{e}_{1,2}$ corresponding to the electric field. In the laser system $e_1^{(a)} = e_2^{(a)}$ [27] and $\delta a_1 = \delta a_2 = \delta a$. If $|\delta a| \ll a_{1,2}$ the expression (A2) takes the form

$$a(t) \approx \exp(i\phi) \left[ a_1 \exp(i\Delta\phi_1/2) \exp(-i\omega_1 t) + a_2 \exp(-i\Delta\phi_2/2) \exp(-i\omega_2 t) \right] \quad (A3)$$

where $\Delta\phi_{1,2} = \Delta\phi \pm \frac{2|\delta a|}{a_{1,2}} \sin\left( \arg(\delta a) - \phi \mp \Delta\phi/2 \right)$.

If $a_1 = a_2$ and $\arg\langle \mathbf{e}_2 | \mathbf{e}_1 \rangle \approx 2\pi n, n = 0, 1, ...$, then $\Delta\phi_1 = \Delta\phi_2$ and the spontaneous emission leads only to a change in the relative phase ($\Delta\phi \to \Delta\phi_1$) while the common phase $\phi$ is unchanged. In an SCL, the first condition ($a_1 \approx a_2$) holds true at all moments in time. The second condition is satisfied at the maxima of the electromagnetic field; this makes the main contribution to the laser radiation and determines the linewidths.

**Funding**

Russian Science Foundation (No. 20-72-10057).

**Acknowledgments**

The study was supported by a grant from Russian Science Foundation (project No. 20-72-10057).I.V.D., Z.A.A. and E.S.A. thank foundation for the advancement of theoretical and mathematics"Basis".